# 43 W, 1.55 µm and 12.5 W, 3.1 µm dual-beam, sub-10 cycle, 100 kHz optical parametric chirped pulse amplifier

Mark Mero,[1,*] Zsuzsanna Heiner,[2,3] Valentin Petrov,[1] Horst Rottke,[1] Federico Branchi,[1] Gabrielle M. Thomas,[1] and Marc J. J. Vrakking[1]

[1]Max Born Institute for Nonlinear Optics and Short Pulse Spectroscopy, 12489 Berlin, Germany
[2]School of Analytical Sciences Adlershof SALSA, Humboldt-Universität zu Berlin, 12489 Berlin, Germany
[3]Department of Chemistry, Humboldt-Universität zu Berlin, Berlin 12489, Germany
*Corresponding author: mero@mbi-berlin.de

**We present a 100 kHz optical parametric chirped pulse amplifier (OPCPA) developed for strong-field attosecond physics and soft-X-ray transient absorption experiments. The system relies on noncollinear potassium titanyl arsenate (KTA) booster OPCPAs and is pumped by a 240 W, 1.1 ps Yb:YAG Innoslab chirped pulse laser amplifier. Two optically synchronized infrared output beams are simultaneously available: a 430 µJ, 51 fs, carrier-envelope phase (CEP) stable beam at 1.55 µm and an angular-dispersion-compensated, 125 µJ, 73 fs beam at 3.1 µm.**

High-repetition-rate (i.e., >> 10 kHz) ultrafast laser sources operating in the near-infrared (NIR, 0.8-3 µm) and the mid-infrared (MIR, 3-30 µm) combined with state-of-the-art charged-particle detection systems enable the investigation of strong-field physics in hitherto unexplored regimes [1]. Long-wavelength drivers can also be exploited to push the cutoff photon energy in high-harmonic generation (HHG) towards, and into the soft-X-ray regime (SXR, > 120 eV) [2]. For the study of biomolecules in their natural aqueous environment, the so-called water window (284-543 eV) constitutes a particularly interesting spectral range [2]. However, the unfavorable scaling of the single-atom response in HHG, $\lambda^{-(5-6)}$, can only partly be overcome by phase-matching schemes [3]. Therefore, for high single-shot SXR photon numbers, optical driver pulse energies at the multi-100 µJ level are required even at center wavelengths below 2 µm, which has so-far limited table-top water-window SXR sources to repetition rates of ≤ 1 kHz [2].

Diode-pumped, Tm-doped fiber chirped pulse laser amplification near 2 µm is a promising, scalable technology for pumping water-window SXR sources. Recently, 252 µJ, sub-50 fs pulses were generated at 1.95 µm at a repetition rate of 61 kHz using Tm-fiber laser technology combined with nonlinear pulse compression [4]. Another viable route to obtain the required pulse parameters in the 1.5-4 µm spectral range is based on optical parametric chirped pulse amplifiers (OPCPAs) driven by diode-pumped ps Nd or Yb lasers. Accordingly, the aforementioned applications led to a recent surge in the development of such OPCPA systems [5-12]. The pulse energy and average power of current state-of-the-art OPCPA systems operating at ~3 µm with repetition rates of ~100 kHz now exceed 100 µJ and 10 W, respectively [10,12]. In order to avoid complications arising from an angularly-dispersed signal or idler beam, these systems rely on bulk, angle-tuned crystals used in the conventional non-collinear amplifier configuration, where the seed pulses are free from angular dispersion and fall in the MIR spectral range. $KNbO_3$-based noncollinear parametric booster amplifiers have enabled a record-high average power of 19 W in carrier-envelope phase (CEP) stable, 118 µJ, 97 fs pulses at 3.25 µm [10]. In another development, bulk $LiNbO_3$ has been successfully employed in noncollinear booster amplifiers in the 15 W, 3.1 µm, 100 kHz OPCPA system at the ELI-ALPS facility, delivering CEP-stable, 150 µJ pulses with a duration of 42 fs [12]. Nevertheless, further power scaling is expected to be hindered by the material properties of $KNbO_3$ and $LiNbO_3$. A more promising material is $KTiOAsO_4$ (KTA), as it exhibits significantly lower OH absorption and higher resistance to laser-induced damage, thermal shock, photodarkening, and photorefractive effects [13-17]. Apart from the lower effective nonlinearity, the main disadvantage of KTA for this particular application is that, when pumped at 1 µm, the conventional noncollinear amplification scheme only yields an improved parametric gain bandwidth in the 1.5-4 µm spectral range when using <2 µm seeding, and at the cost of an angularly-dispersed MIR idler beam. As a result, KTA has only been considered in the wider laser community for use in collinear geometry in the aforementioned spectral range [6,11,13].

In this Letter, we present a 100 kHz ultrafast OPCPA system based on noncollinear KTA booster amplifiers delivering a total average output power in the NIR signal and MIR idler beams of 66 and 55.5 W before and after chirp compensation, respectively. Despite the noncollinear geometry, both signal and idler beams are available for experiments. The generation of high-quality MIR idler pulses was achieved by implementing a simple angular dispersion compensation scheme [18]. The OPCPA was designed for simultaneous delivery of (i) a 1.55 μm, 50 fs, CEP-stable beam for driving, following nonlinear pulse compression, a high-photon-flux HHG source reaching SXR photon energies extending into the water window and (ii) an optically synchronized 3.1 μm idler beam for investigating strong-field attosecond physics.

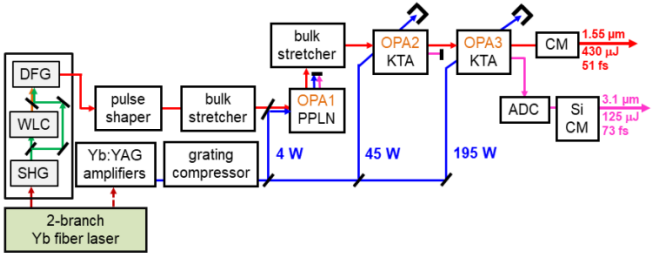

Fig. 1. Scheme of the dual-beam, 100 kHz OPCPA system. SHG, second-harmonic generation; WLC, white-light continuum generation; DFG, difference-frequency generation; OPA, optical parametric amplifier; PPLN, periodically poled LiNbO$_3$; KTA, KTiOAsO$_4$; CM, chirped mirrors; ADC, angular dispersion compensation; Si, silicon window.

Figure 1 shows the conceptual scheme of our OPCPA. Compared to our previous system [6,9], various upgrades were implemented and a final parametric booster amplifier stage was added. One of the upgrades involved replacement of the two-branch, 80 MHz, 1.56 μm Er-fiber laser seeder with a two-branch, 1.03 μm, 100 kHz Yb-fiber laser. This way, the Yb:YAG amplifiers are seeded by pulses from a robust, high-stability Yb-fiber oscillator-amplifier system, instead of a supercontinuum module pumped by 1.56 μm pulses. The first branch of the Yb-fiber laser delivers few-10 nJ, stretched seed pulses *via* fiber to the input port of the Yb:YAG amplifier system. The second branch is a free-space beam of 20 μJ, 290 fs pulses that drive a home-built difference frequency generation (DFG) unit supplanting the earlier Er-fiber master oscillator-amplifier module. Guided by the scheme described in [19], the DFG unit is based on a 2-mm-long, type-I beta barium borate (BBO) crystal and delivers 180 nJ, 1.55 μm, passively CEP-stabilized pulses for seeding the OPCPA chain, with a transform-limited duration of 39 fs. Compared to the Er-fiber laser seeder, the DFG unit provides a 50-fold increase in seed pulse energy. In addition, it offers CEP stability, which previously could not be attained even when using the feed-forward CEP-stabilization technique [20]. In this case, active CEP-stabilization was prevented by the very high, correlated amplitude and frequency noise in the carrier-envelope beat note, possibly due to the saturable absorber mirror of the Er-laser operating on a narrowband excitonic resonance [21].

In another upgrade implemented in the OPCPA system, the transmission grating-based compressor of the Yb:YAG amplifier was replaced by a compressor based on dielectric reflection-type gratings, leading to an increase in the throughput of the compressor from 65% to 80%. At the same time, the power of the main Yb:YAG Innoslab amplifier was reduced from 440 to 300 W to extend the lifetime of critical optical components. After chirp compensation, 240 W, 1.1 ps pulses are available to pump the OPCPA chain.

The OPCPA chain consists of three stages. After exiting a 640-pixel spatial light modulator-based pulse shaper, the 1.55 μm OPCPA seed pulses are stretched by 94 mm of SF57 glass to 430 fs and are combined with the pump pulses at a dichroic mirror. As in our previous system, the first stage (OPA1) is collinear and based on an anti-reflection (AR) coated, 2-mm-long 5%-MgO-doped, fan-out periodically poled lithium niobate (PPLN) crystal [6]. In the new system, using an unchanged pump pulse energy of 40 μJ, the focusing conditions were adapted to the higher OPCPA seed pulse energy. The seed pulse energy incident on the PPLN crystal is 50 nJ, and is amplified to 3.3 μJ by the first stage. The signal pulses are then further stretched by 47 mm of SF57 to 600 fs and seed the second stage (OPA2), which is identical to that in our the previous system [6]. This stage is based on an AR-coated, 4-mm-long, type-II KTA crystal operated in the noncollinear amplifying arrangement in the X-Z plane, and delivers up to 100 μJ signal pulse energy, when pumped at an energy of 450 μJ. At this stage, up to 50 μJ, 3.1 μm idler pulses are generated in an angularly-dispersed beam. In [9], we explored the feasibility of angular dispersion compensation of this idler beam using an AR-coated silicon prism, which led to significant astigmatism and laterally varying chirp in the beam and the conclusion that a design based on reflection gratings is necessary given our broad spectral bandwidth. In the current system, the idler output beam from OPA2 is dumped.

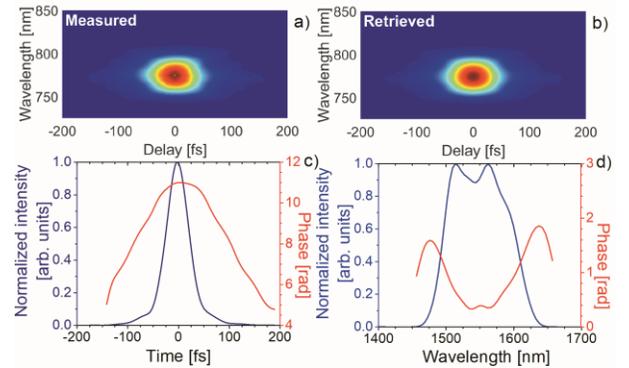

Fig. 2. (a) Measured and (b) retrieved SHG-FROG traces for the chirp-compensated, 430 μJ signal pulses. Reconstructed temporal (c) and spectral (d) intensity and phase. The retrieved pulse duration is 51 fs.

The third and final OPCPA stage (OPA3) is an AR-coated, 2-mm-long, type-II KTA crystal that is once more used in noncollinear geometry. It is pumped by 1.9 mJ pulses, which were down-collimated to a radius of $w_{1/e^2} = 1.25$ mm, leading to approximately the same peak pump intensity as in the second stage (i.e., 70 GW/cm$^2$) and a gain of ~5. In the final OPCPA stage, the signal pulse energy is boosted to 470 μJ without any loss in spectral bandwidth and, simultaneously, angularly-dispersed idler pulses with an energy of 190 μJ are generated. Chirp compensation for the 1.55 μm beam was achieved using 24 reflections off high-dispersive chirped mirrors (-300 fs$^2$/reflection) and manual fine tuning of the pulse shaper. The throughput of the chirped mirror compressor is 91% leading to 430 µJ (i.e., 43 W) nearly transform-limited pulses. Figure 2 shows measured and reconstructed second-harmonic generation frequency-resolved optical gating (SHG-FROG) traces together with the retrieved temporal and spectral profiles. The measured signal pulse duration is 51 fs, corresponding to ≲ 10 optical cycles, while the transform-limited value calculated from the spectrum is 47 fs.

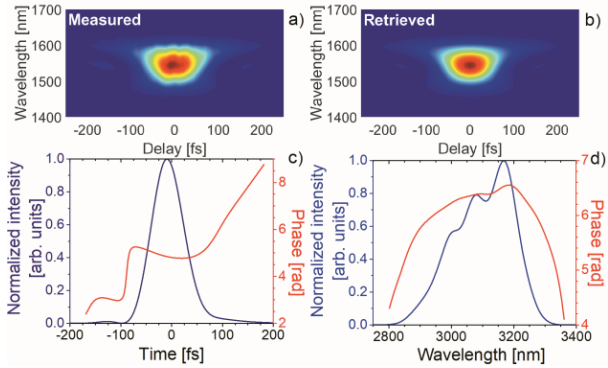

Fig. 3. (a) Measured and (b) retrieved SHG-FROG traces for the chirp-compensated, 125 µJ idler pulses. Reconstructed temporal (c) and spectral (d) intensity and phase. The retrieved pulse duration is 73 fs.

Since the noncollinear amplifying geometry comes at the cost of an angularly-dispersed idler beam, proper compensation is required to obtain a beam that can be focused to high intensities. We implemented a simple, reflection grating-based setup that was first carefully tested with respect to the achievable beam quality on a small-scale infrared OPA system relying essentially on the same type of noncollinear KTA booster amplifier as our final OPCPA stage [18]. We found that by employing a grating on a copper substrate, the setup is scalable to high average powers. For the design and implementation, we followed the recipe described in [18]. Briefly, the external (i.e., in air), ~120 µrad/nm angular dispersion of the idler pulses generated in OPA3 is cancelled by imaging the crystal plane onto a ruled grating (40.96 lines/mm, 3.2 µm blaze wavelength) at a transverse magnification of 3 and an angular demagnification of 1/3 (i.e., 120 µrad/nm / 3 ≈ 41 µrad/nm), using a single spherical reflector. The measured diffraction efficiency of the grating into the -1st order, a few degrees off the Littrow configuration, is 77%. For rough alignment, the distances were initially chosen according to the Gaussian thin lens equation and were optimized by minimizing the spatial chirp in the far field. The temporal chirp of the corrected idler beam is compensated by transmission through a 10-mm-long AR-coated silicon window and 8 reflections off high-dispersive chirped mirrors (+500 fs$^2$/reflection). The total throughput of the angular dispersion and chirp compensation units is 66%, leading to a pulse energy of 125 µJ (i.e., 12.5 W). Figure 3 shows measured and reconstructed SHG-FROG traces together with the retrieved temporal and spectral profiles. The measured idler pulse duration is 73 fs corresponding to ~7 optical cycles, while the transform-limited value is 71 fs. We note that this MIR temporal profile is obtained under conditions, where the compression of the signal pulses is optimized using the pulse shaper resulting in some uncompensated high-order idler chirp. However, the temporal broadening effect of this residual idler chirp is negligible. By fine tuning the chirp of the signal pulses using the pulse shaper, it is possible to reduce this high-order residual chirp in a straightforward way.

Figure 4 shows the near-field profiles of the 43 W signal and the 12.5 W idler beams, 400 mm behind the respective chirped mirror compressors. Both beams follow a nearly Gaussian spatial distribution with full-width-at-half-maximum (FWHM) diameters of 2-2.5 mm.

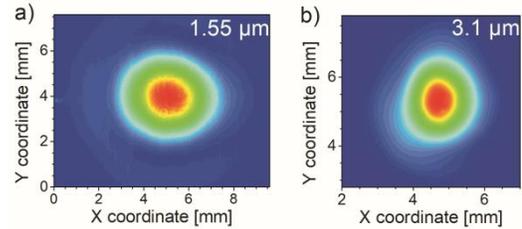

Fig. 4. Near-field profiles of the chirp-compensated 430 µJ signal (a) and 125 µJ idler (b) beams, measured 400 mm behind the respective chirped mirror compressors.

The open-loop CEP stability of the 1.55 µm signal beam at the 9-W-level was characterized by f-2f interferometry using a single-beam setup following the procedure we used in [22]. Briefly, a multi-octave, single-filament supercontinuum was generated by focusing a small fraction of the signal pulses into a 4-mm-thick uncoated YAG window. The continuum was collimated and focused into a 0.3-mm-long, type-I BBO crystal, which was angle tuned to frequency double the spectral range at 1.54 µm. The orthogonally polarized fundamental and second-harmonic pulses at 0.77 µm were relay-imaged onto the slit of a CCD spectrometer. The visibility of the spectral fringes was optimized using a wire-grid polarizer. Each spectrum was averaged over 1200 pulses (i.e., 12 ms integration time). A typical spectrum and the extracted CEP fluctuations as a function of time are shown in Fig. 5. The corresponding root mean square (RMS) CEP jitter was 390 mrad over a 10 second period dominated by low-frequency noise, which can be compensated by an active CEP stabilization unit in a straightforward way.

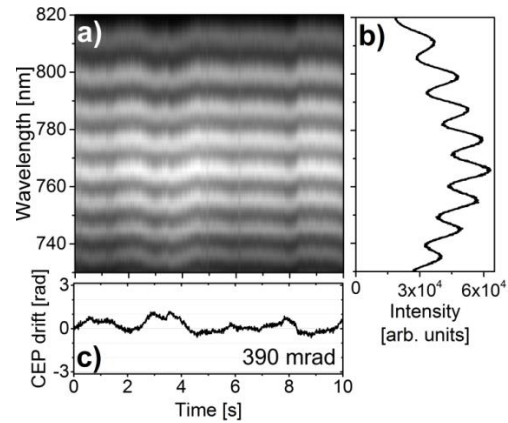

**Fig. 5.** Open-loop CEP measurements. (a) f-2f spectra as a function of time, (b) typical f-2f spectrum, and (c) time evolution of the CEP offset characterized by an RMS value of 390 mrad.

The long-term RMS average power stability of both beams is below 0.5% when measured using a power meter. Long-term stability is further corroborated by strong-field ionization experiments using 8 W of the 1.55 µm signal beam and a reaction microscope as particle detector. Figure 6 shows the multi-photon ionization rate of Ar atoms as a function of time over a period of 1.4 hours. The data points were taken with an integration time of 1 s. The data follow a normal distribution with a mean of 2566 Hz and a standard deviation of 95 Hz. As the standard deviation due to shot noise is 51 Hz, the remaining fluctuations correspond to a standard deviation of 80 Hz, or 3.1%. Although a quantitative characterization of the intensity variations at focus is not possible from this measurement, the low measured fluctuations in the 20-photon ionization rate of Ar (i.e., ionization potential = 15.76 eV) are a testament to the average power stability of our OPCPA system.

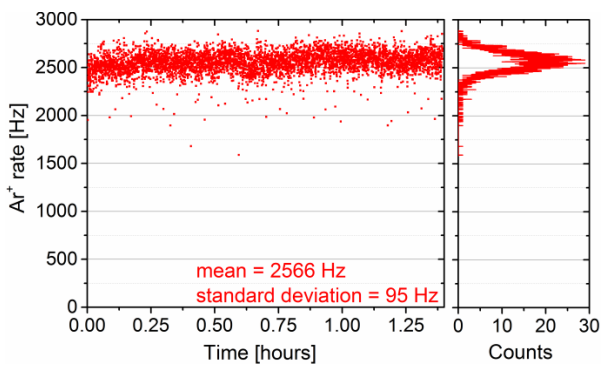

Fig. 6. Rate of Ar+ generation by focused 1.55 µm signal pulses, measured using a reaction microscope. Each data point corresponds to a measurement taken at an integration time of 1 s. The background rate was negligible.

In conclusion, we have demonstrated a dual-beam 100 kHz infrared OPCPA system delivering an unprecedented average power at 1.55 µm in 430 µJ, 51 fs, passively CEP-stabilized pulses together with 125 µJ, 73 fs pulses at 3.1 µm. In contrast to existing OPCPA systems utilizing noncollinear $KNbO_3$ or $LiNbO_3$ parametric booster amplifier stages that are seeded in the MIR, our system is based on noncollinear KTA booster amplifiers seeded at 1.55 µm. By employing a simple angular dispersion compensation technique, we showed that despite the noncollinear amplifying geometry, KTA can be efficiently used for generating broadband, high-quality MIR pulses at high average powers. Further development will include the implementation of active CEP stabilization and nonlinear pulse compression of the 1.55 µm beam. The 1.55 µm output of this unique system will serve as the pump of a high-flux soft-X-ray source with a spectrum reaching the water window, while the 3.1 µm beam will provide optically synchronized driver pulses for studying strong-field processes.

**Funding.** Leibniz-Gemeinschaft (SAW-2012-MBI-2); The European Union's Horizon 2020 research and innovation programme (654148); Deutsche Forschungsgemeinschaft (DFG) (GSC 1013 SALSA).

**Acknowledgement.** Z.H. acknowledges funding by a Julia Lermontova Fellowship from DFG, No. GSC 1013 SALSA.